Deep-dGFP: Deep Learning Approach for Large-Scale, Real-Time Quantification of Green Fluorescent Protein-Labeled Biological Samples in Microreactors

Yuanyuan Wei[1], Sai Mu Dalike Abaxi [1], Nawaz Mehmood[1], Luoquan Li[1], Fuyang Qu[1], Guangyao Cheng[1], Dehua Hu[1], Yi-Ping Ho[1, 2, 3, 4], Scott Wu Yuan[1, *], Ho-Pui Ho[1, *]

[1] Department of Biomedical Engineering, The Chinese University of Hong Kong, Shatin, Hong Kong SAR, China. E-mail: wyuan@cuhk.edu.hk; aaron.ho@cuhk.edu.hk

[2] Centre for Biomaterials, The Chinese University of Hong Kong, Hong Kong SAR, China

[3] Hong Kong Branch of CAS Center for Excellence in Animal Evolution and Genetics, Hong Kong SAR, China

[4] The Ministry of Education Key Laboratory of Regeneration Medicine, Hong Kong SAR, China

[*]Correspondence: wyuan@cuhk.edu.hk;.aaron.ho@cuhk.edu.hk.

**Abstract**

Absolute quantification of biological samples entails determining expression levels in precise numerical copies, offering enhanced accuracy and superior performance for rare templates. However, existing methodologies suffer from significant limitations: flow cytometers are both costly and intricate, while fluorescence imaging relying on software tools or manual counting is time-consuming and prone to inaccuracies. In this study, we have devised a comprehensive deep-learning-enabled pipeline that enables the automated segmentation and classification of GFP (green fluorescent protein)-labeled microreactors, facilitating real-time absolute quantification. Our findings demonstrate the efficacy of this technique in accurately predicting the sizes and occupancy status of microreactors using standard laboratory fluorescence microscopes, thereby providing precise measurements of template concentrations. Notably, our approach exhibits an analysis speed of quantifying over 2,000 microreactors (across 10 images) within remarkably 2.5 seconds, and a dynamic range spanning from 56.52 to 1569.43 copies $\mu L^{-1}$. Furthermore, our Deep-dGFP algorithm showcases remarkable generalization capabilities, as it can be directly applied to various GFP-labeling scenarios, including droplet-based, microwell-based, and agarose-based biological applications. To the best of our knowledge, this represents the first successful implementation of an all-in-one image analysis algorithm in droplet digital PCR (polymerase chain reaction), microwell digital PCR, droplet single-cell sequencing, agarose digital PCR, and bacterial quantification, without necessitating any transfer learning steps, modifications, or retraining procedures. We firmly believe that our Deep-dGFP technique will be readily embraced by biomedical laboratories and holds potential for further development in related clinical applications.

## 1. Introduction

The precise determination of molar concentration or number concentration of particles in a defined volume is both important and challenging, especially when the purity of the sample is uncertain [1], [2]. However, most methods rely on standard samples (calibrators) that are often difficult to obtain [2]. Additionally, if the sample and standard differ in their characteristics (e.g., amplification efficiency in the standard-curve method of real-time quantitative polymerase chain reaction), quantification errors of significant magnitude can occur. Considering the significance of particle diversity and heterogeneity in quantification, single-particle counting for absolute quantification has gained considerable importance. For instance, when quantifying the number of target genes in a community sample, droplet digital PCR (ddPCR) has emerged as a highly sensitive, precise, and accurate method in the scientific community only in the last decade or so [3]–[6]. It represents one of the most sensitive, precise, and accurate methods for amplifying and quantifying sequences of various nucleic acids (DNA, RNA, and cDNA). The typical ddPCR workflow consists of four steps: sample dilution, sample partitioning, thermal cycling, and data analysis. By analyzing the proportions of positive droplets (fluorescent droplets) and negative droplets (exhibiting little to no fluorescence) with Poisson statistics, the concentration of the DNA template in the original sample can be determined [7], [8]. However, the accurate identification of positive droplets in ddPCR images is critical to maintaining target nucleic acid quantitative analysis precision. Regarding droplet analysis, the conventional approach involves analyzing fluorescence images of stationary droplets in a microchamber or on a glass slide. Current approaches for ddPCR image analysis rely on software tools with manual threshold segmentation [9], [10], such as Image J [11], Fiji [12], and CellProfiler [13], which can be labor-intensive and lack scalability. While alternative approaches like in-flow interrogation [14] of droplet fluorescence have been proposed, they often require expensive and complex devices, limiting the speed, portability, and operational simplicity of ddPCR.

Deep learning has emerged as a powerful tool in biosensing and data analysis, including analyzing ddPCR images [15]. It offers significant advantages in handling noisy and low-resolution sensing data that often have overlapping features. By employing techniques such as categorization, anomaly detection, noise reduction, object identification, and pattern recognition [16], [17], deep learning is capable to reveal hidden relationships between sample parameters and sensing signals. One specific application is the use of deep learning, such as the Mask Region Convolutional Neural Network (Mask R-CNN) method, to simplify droplet reading and analysis [18]. While this method has shown superior performance compared to other image analysis algorithms, its quantification capabilities for droplets are still relatively unexplored, currently limited to detecting positive compartments. Further research is needed to fully explore the potential of deep-learning technique for absolute droplet quantification. Additionally, conventional machine vision algorithms struggle to extract all relevant information from complex backgrounds, and two-stage methods that rely on grayscale values for droplet localization and classification require high-quality imaging [19]. However, these reports indicate the potential for deep-learning to be applied to various similar quantification tasks involving fluorescence-probe (e.g., TaqMan™ probe, or DNA-binding dye EvaGreen® and SYBR Green) labeled biological samples.

Considering the fluorescence-based probes, green fluorescence protein (GFP) is widely used not only in the ddPCR process but also in many other cell and molecular biology experiments as a reporter of expression [20]–[22]. Unlike most small fluorescent molecules like FITC (fluorescein isothiocyanate), which can be highly phototoxic in live cells, fluorescent proteins such as GFP are usually much less harmful when illuminated. One major advantage of GFP is its heritability and noninvasive visualization, requiring only blue/violet/ultraviolet illumination and emitting green light. For instance, SYBR Green can bind to any double-stranded DNA sequence and endorses fluorescence emitted from non-specific qPCR products, such as primer dimers. This allows both positive and negative microreactors to exhibit fluorescence compared to their background as a reference, while the intensities differ. By analyzing the intensity difference of sample-encapsulated microreactors, our trained deep-learning algorithm can automatically quantify biological information, including number, diameter, concentration, expression, and mutation.

Here, to provide a powerful yet mobile and inexpensive tool for biological template absolute quantification, we introduce a generalized deep-learning-enabled pipeline for GFP-labeled biological samples. After GFP-labeled samples are partitioned into microreactors, the developed deep-learning algorithm is capable of automatically segmentation, classifying, and analyzing captured fluorescence images in a real-time manner. The analysis results are real-time displayed on our developed GUI (graphical user interface), with a very large dynamic range from 56.52 to 1569.43 copies $\mu L^{-1}$. To validate its generalization, this analysis technique has been applied to droplet digital PCR, droplet single-cell sequencing, microwell digital PCR, and other biological applications. This Deep-dGFP algorithm represents a cost-effective and automated analysis tool that constitutes a significant advancement in the biological quantification field, due to the integration of accessible hardware (microfluidics chip, thermal cycler, and camera-coupled fluorescence microscope) with a state-of-the-art deep learning algorithm. Compared to conventional analysis method that requires a flow-cytometer or human counting, this technique is significantly more robust, timesaving, simple to operate, and extremely cost-effective.

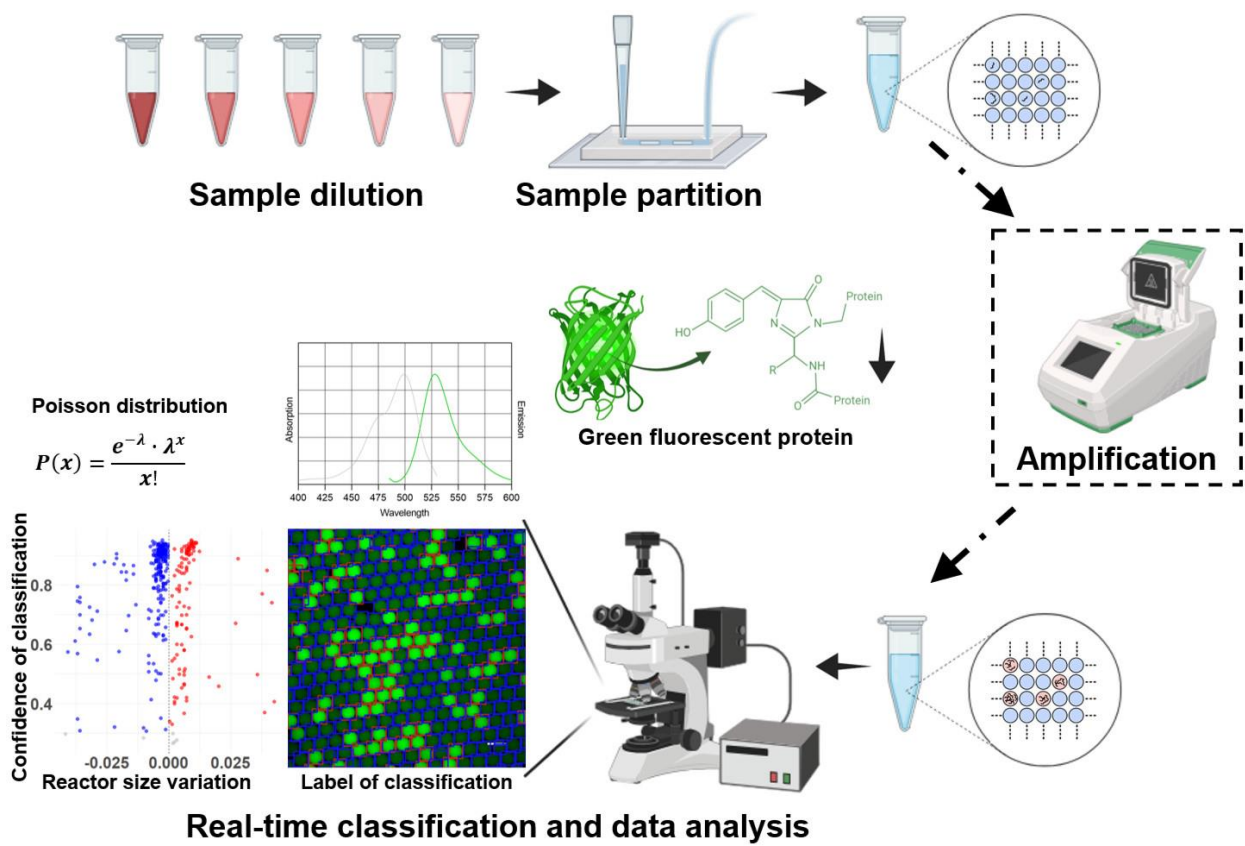

**Fig. 1 Schematic of the Deep-dGFP: deep learning enabled pipeline for automatic real-time quantification of biological samples labeled with green fluorescence protein (GFP).** The process involves diluting and partitioning the target templates into over 20,000 microreactors. Subsequently, around 2,000 microreactors are imaged using fluorescence microscope FITC (Fluorescein isothiocyanate) mode. These captured images are displayed in real-time on a graphical user interface (GUI) with classification labels and analyzed using a deep-learning algorithm. Microreactors containing templates are classified as positive (labeled in red), while those without templates are classified as negative (labeled in blue). The analysis is performed by accumulating results from sequential frames captured at a rate of 1 frame per second, achieved by moving the sample stage of the fluorescence microscope. The inferenced target concentration of the original sample is determined using positive and negative fluorescence data from the sample droplets, as well as data fitting to a Poisson distribution. Both positive and negative microreactors exhibit fluorescence compared to their background, although the intensities differ. The GFP chromophore emits green fluorescence upon absorbing blue light and consists of just three amino acids in a barrel shape. Based on the fluorescence intensity difference of sample-encapsulated microreactors, the trained deep-learning algorithm is capable of quantifying biological information including number, diameter, concentration, expression, and mutation automatically. If amplification of templates is required, PCR (Polymerase chain reaction) cycling for template amplification is necessary.

2. **Results**

We validated our Deep-dGFP technique by performing automatic and real-time quantification analysis of ddPCR experiments. As illustrated in **Fig. 1**, we diluted the glyceraldehyde-3-phosphate dehydrogenase (GAPDH) sample into serial concentrations, ranging from the no template control (NTC) to 5 ng/μL (including 100 fg/μL, 1 pg/μL, 10 pg/μL, 100 pg/μL, and 5 ng/μL). Using a homemade microfluidic chip and a customized lab-on-a-chip actuation system, we partitioned the serially diluted templates and PCR reagents into over 20,000 monodispersed droplets, each sized at 46.37 ± 1.64 μm (correlated volume of 52.20 pL). Following thermal cycling, we introduced SYBR Green dye into each PCR stripe tube. The droplets, dispensed into microchamber chips for stationary, were then captured using FITC mode fluorescence imaging for generating dataset. These real-time captured images, along with classification labels, were displayed on our graphical user interface (GUI) through a laptop connected to the microscope via an Application Programming Interface (API) cable (as shown in **Fig. 2(g)**). Simultaneously, labeled images and analyzed results, including class, confidence, coordinate of each droplet, and dimensions of correlated bounding box, were output and saved into a designated folder. The analysis results of the droplets were plotted, with the horizontal axis representing droplet size and the vertical axis representing classification confidence (refer to **Supplementary Fig. 1**). In the plot, droplets containing templates were classified as positive and colored in red, while droplets without templates were classified as negative and colored in blue. Additionally, dots with a classification confidence smaller than 0.3 were labeled grey. The plotting results were accumulated from different frames captured at a rate of 1 frame per second, achieved by moving the sample stage of the fluorescence microscope, resulting in a total of over 2,000 droplets. The concentration of the original sample was determined by fitting the data to a Poisson distribution. Further details of the ddPCR experiments can be found in the Materials and Methods section.

**Figure 2** illustrates the workflow and functional performance of our automated image analysis algorithm. The labeling and training procedures consisted of five steps: (1) Image labeling, (2) Data augmentation, (3) Data splitting, (4) Training the detection model (Yolo-V5m), and (5) Analyzing the detection results. Both frames and videos were utilized for data analysis during the detection process. The image labeling process involved capturing, screening, and subjecting fluorescence imaging video frames, as shown in **Fig. 2(a)**. The droplets were accurately labeled and categorized into two classes: the "positive" class (represented by the color red) or the "negative" class (represented by the color blue). This categorization was achieved through the manual initiation of bounding boxes and subsequent visual examination. The analysis pipeline incorporated both area and labeling information during the training process as **Fig. 2(b)** shows. Further details regarding the implementation and training of the model can be found in the Materials and Methods section. Following the completion of training, droplet segmentation and classification were automatically performed, with **Fig. 2 (c)** providing a graphical representation of the output. The analysis results of serial-diluted ddPCR were plotted in **Fig. 2 (d)**. Sample concentrations were automatically calculated based on the Poisson distribution. The statistical analysis of droplet classification fluorescence intensity reveals a significant correlation. To assess the reliability of the YOLOv5m model, we employed differentiated evaluation metrics to scrutinize the datasets, as depicted in **Fig. 2 (e)** and **(f)**. **Fig. 2(e)** presents the visualization of quantitative results, including mean average precision (mAP), intersection over union (IoU), and the precision-recall curve. Impressively, the trained YOLOv5m model demonstrated remarkable accuracy, achieving an 88% mAP with IoU values of up to 0.8. This ensures the reliability and robustness of the developed algorithm for analyzing fluorescence images of GFP-labeled micro-reactors. The precision-recall curve in **Fig. 2(f)** follows a similar trend, with a moderate to high (0.5-0.8) IoU threshold yielding both high precision and recall values.

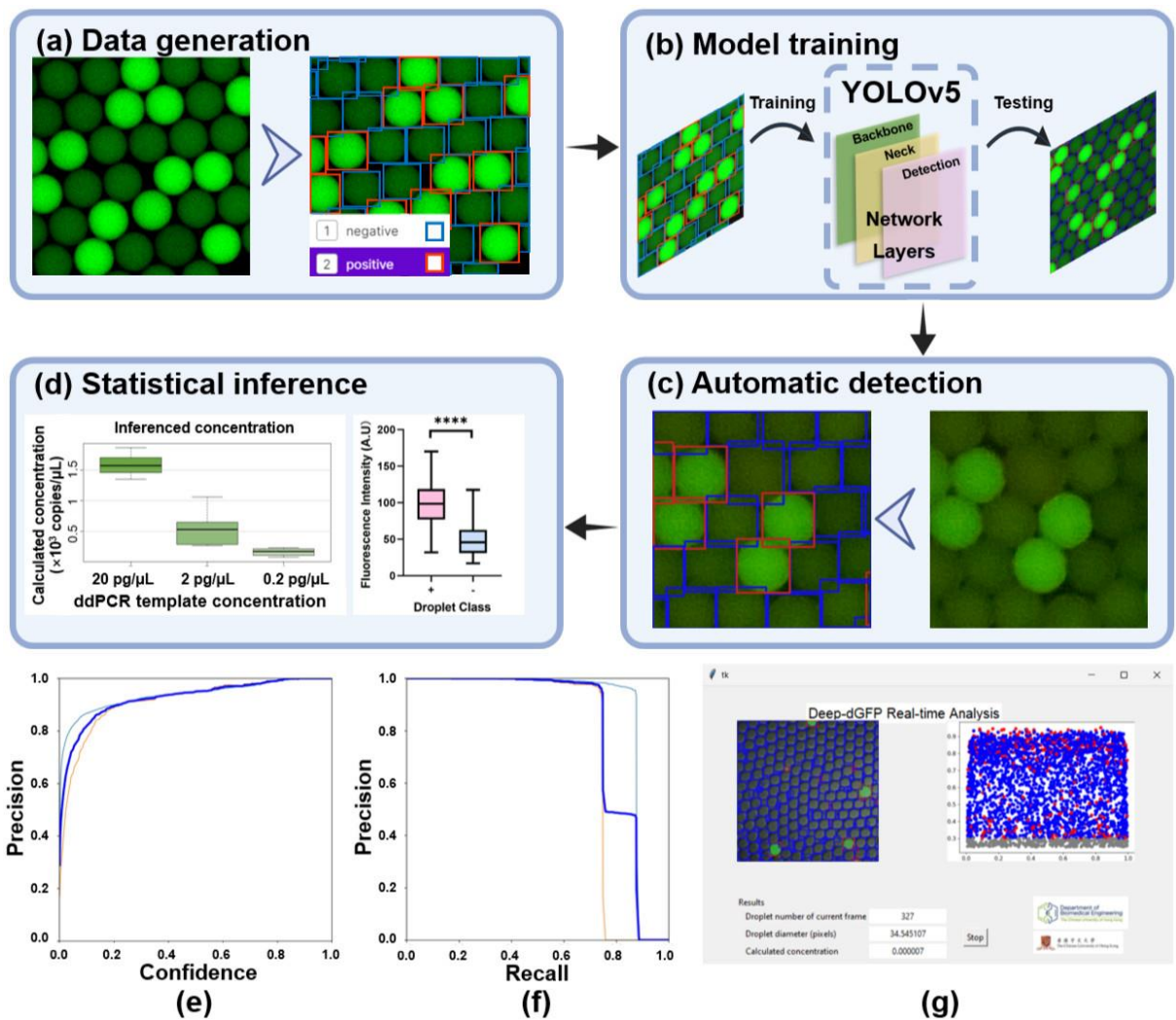

**Fig. 2 Illustration of the workflow and performance of our deep learning-enabled real-time image analysis pipeline.** (a) The droplets in the training dataset are manually labeled into positive (red) and negative (blue) classes. (b) The Yolo-V5m deep learning algorithm is trained with an 8:1:1 ratio for the training, validation, and testing datasets, respectively. (c) Automatic droplet segmentation and classification of input images or video streams captured by a fluorescence microscope can be realized in a real-time manner. (d) By fitting the analysis results into the Poisson distribution, sample concentrations are automatically calculated. To demonstrate the reliability of our analysis results, statistical analysis of droplets' fluorescence intensity and classification reveals a significant correlation. (e) The trained Yolo-V5m algorithm exhibits remarkable accuracy, achieving an 88% mean average precision (mAP) with an intersection over union (IoU) threshold of up to 0.8, indicating robust performance. However, as IoU increases to 0.9 (the predicted box had to overlap with the true box for more than 90%), the mAP decreases due to the more stringent

requirements. (f) The precision-recall curve follows a similar trend, with a moderate to high (0.5-0.8) IoU threshold yielding both high precision and recall values. (g) GUI for Deep-dGFP real-time microreactors segmentation and classification, and data analysis. The plotting results are accumulated from different frames captured (1 frame per second) by moving the sample stage of the fluorescence microscope. The inferenced concentration of the original sample relies on two elements: positive and negative fluorescence data from the sample droplets and data fitting to a Poisson distribution. Besides the real-time modes by reading from the API (Application Programming Interface) or SDK (Software Development Kits) cable of the fluorescence microscope, this GUI also enables offline mode by reading the pre-saved images from folders.

We validated our Deep-dGFP approach by analyzing additional ddPCR experiment results conducted in our laboratory. Merged micrographs, derived from bright field and FITC mode fluorescence images via a laboratory fluorescence microscope, were subjected to automated analysis using our pre-trained algorithm. This encompassed droplet segmentation, classification, and frame-by-frame correlated plotting, as demonstrated in **Fig. 3**. The results were plotted against droplet size distribution (x-axis) and classification confidence (y-axis), with droplet diameters normalize according to the ratio of measured droplet diameter to input image side length. In this study, we employed the extracted seahorse (*Hippocampus kuda*) genome as the template, with targeted region located at the *cytochrome c oxidase subunit I (COI)* with an amplicon size of 206 bp. The template was serially diluted into 40 pg, 4 pg, and 0.4 pg per 20 µL PCR system, respectively. As sample dilution resulted in lower concentrations, the number of droplets labeled positive class decreased. Over 2,000 droplets were analyzed for each concentration, with the inferenced concentration detailed in **Fig. 2 (d)**. Accurate segmentation and classification of droplets are essential prerequisites for the successful implementation of deep-ddPCR. Deep-dGFP priories advantages in offering intuitive, quantitative results, informing on droplet size, classification, and class distribution dealing with large-scale experimental images. The validation outcomes underscore the precision and effectiveness of deep-ddPCR, emphasizing its utility in the sensitive detection and accurate quantification of target molecules.

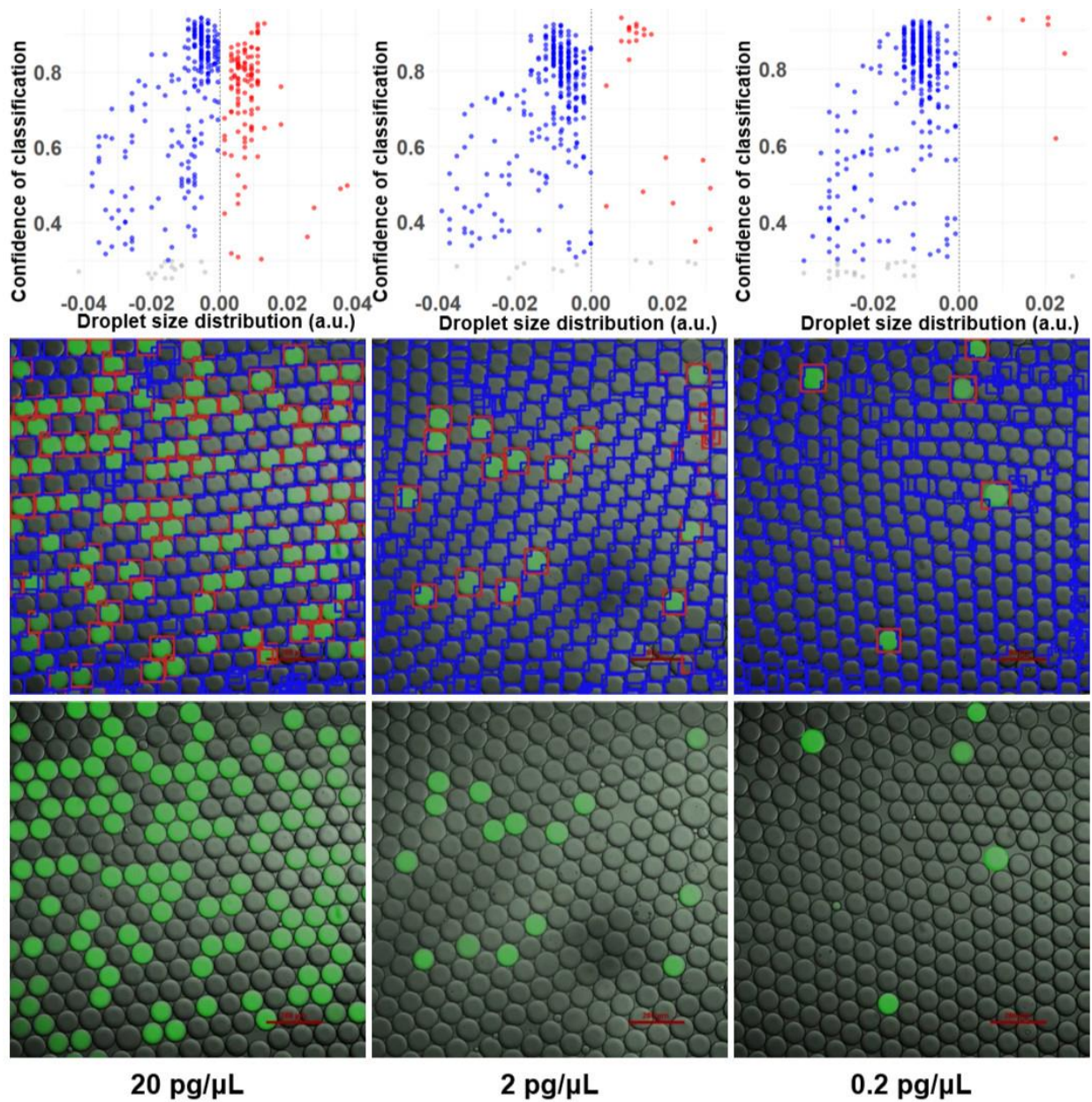

**Fig. 3 The droplet segmentation and classification results of image analysis using representative droplet digital PCR micrographs with serially diluted samples.** Micrographs of the merged images, obtained by capturing bright field and FITC mode fluorescence images using a lab fluorescence microscope, were automatically analyzed. The analysis included droplet segmentation, classification, and correlated plotting of each frame. The analysis results are plotted based on droplet size distribution (horizontal axis) and classification confidence (vertical axis). The droplet diameter is normalized based on the ratio of the measured droplet diameter to the side length of the input image. Droplets with classification confidence below 0.3 are labeled grey. As the sample being diluted to a lower concentration, fewer droplets were labeled and plotted as positive. For each concentration, more than 2,000 droplets were analyzed, with inferenced concentration plotted in **Fig. 2 (d)**. The inter Deep-dGFP, in comparison to fluorescence images with limited information, offers intuitive and quantitative results, providing information about the

size and classification of each droplet, as well as the distribution of different sizes or classes on a large scale. (Scale bar = 200 μm).

In addition to the aforementioned analysis capability of FITC fluorescence images, our Deep-dGFP technique also shows strong generalization capability on images captured under different modes. **Figure 4** illustrates that the trained algorithm can be directly applied to both fluorescence fields and merged fields. The results obtained from FITC fluorescence images exhibit less variation in size, a lower false accept rate (FAR) for classifying positive droplets, and a higher confidence in droplet classification compared to the merged images. However, the merged images yield a lower false rejection rate (FRR) for droplet recognition. We have also applied the algorithm into detecting droplets of different amplification conditions, which results in a difference in fluorescence intensities. With the microreactors being same, PCR cycling was conducted for 25 cycles and 45 cycles, respectively (details in Materials and Methods section). In comparison to the results obtained from 25 cycles, the findings from 45 cycles demonstrate a higher number of droplets being identified and classified as positive.

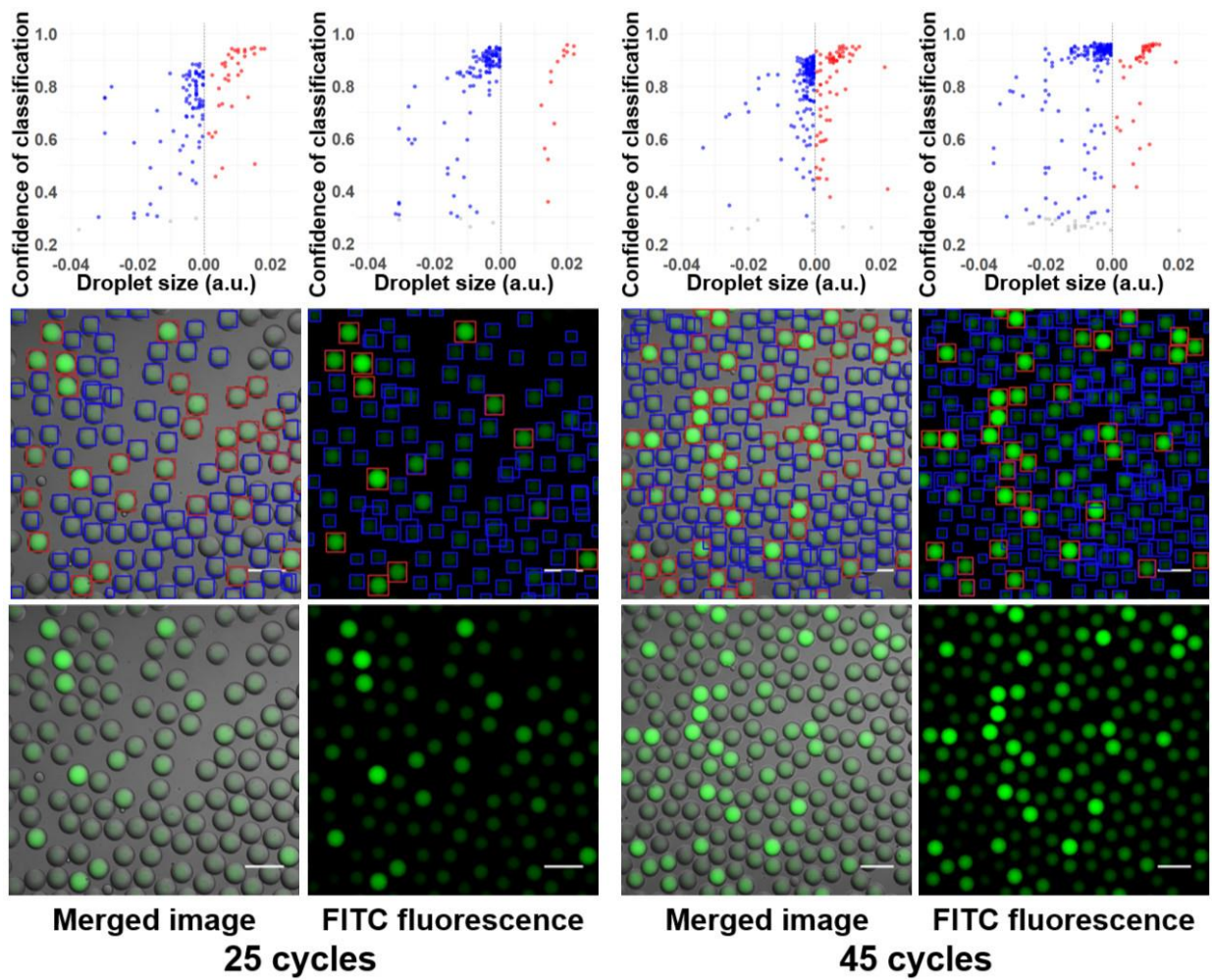

**Fig. 4 The droplet segmentation and classification results of image analysis using representative droplet digital PCR micrographs with different thermal cycling conditions.** Both the FITC fluorescence image and the merged image (combining bright field and fluorescence field images) were analyzed using our developed deep-learning algorithm. In general, the results obtained after 45 cycles showed a higher number of classified positive droplets, compared to the results obtained after 25 cycles. Moreover, the results from FITC fluorescence images exhibited less size variation, a lower false accept rate (FAR) for classified positive droplets, and a higher confidence in droplet classification compared to the results from merged images. However, the merged images showed a lower false rejection rate (FRR) for droplet segmentation. (Scale bar = 100 μm).

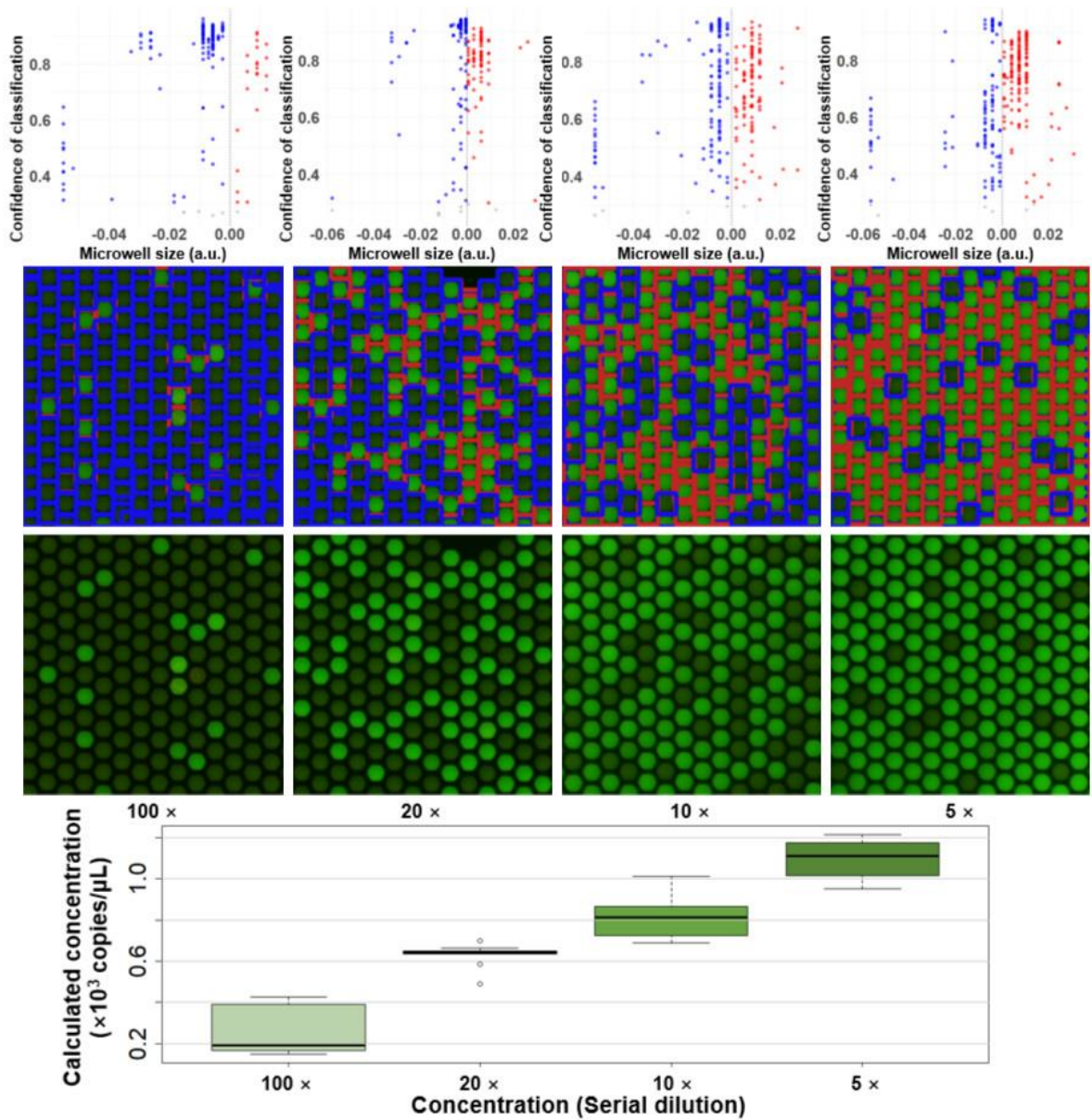

**Fig. 5 Validation of Deep-dGFP generalization capability: the microwell segmentation and classification results of image analysis using representative microwell digital PCR with**

**serially diluted samples.** The developed deep-learning algorithm of Deep-dGFP can be directly applied to microwell-dPCR image analysis tasks without any transfer learning, modifications, or retraining. The results, including microwell segmentation, classification, and corresponding plotting for each frame, are automatically processed by analyzing pre-saved fluorescence images. Fewer microwells were labeled and plotted as positive classes with the sample being diluted to a lower concentration. The detection results encompass a wide dynamic range, spanning template concentrations from $0.258 \times 10^3$ to $1.083 \times 10^3$ copies/μL. For each concentration, more than 2,000 microwells (each reaction-well volume of 755 pL) were analyzed, with inferenced concentration plotted correspondingly. (Scale bar = 100 μm).

Our Deep-dGFP algorithm demonstrates strong generalization capabilities when applied directly to various GFP-labeled microreactor experiments, including droplet-based, microwell-based, and agarose-based applications. Without the need for transfer learning, modifications, or retraining, we successfully analyzed microwell dPCR (performed on 3D Digital PCR chip v2, each reaction-well volume of 755 pL, ThermoFisher Scientific, USA) results of serially diluted concentrations, as shown in **Fig. 5**. The results include correlated segmentation, classification, and plotting of each micrograph, covering a wide dynamic range of template concentrations from $0.258 \times 10^3$ to $1.083 \times 10^3$ copies/μL. In addition to the aforementioned demonstrations, our Deep-dGFP analysis pipeline has also been successfully applied to droplet single-cell sequencing (**Fig. 6 (a)** and **(b)**), agarose digital PCR (**Fig. 6 (c)**), droplet-based digital quantification of bacterial suspension (**Fig. 6 (d)**), and mutation identification of bacteria (**Fig. 6 (e)**). In all these scenarios, microreactors can be automatically and accurately segmented and classified, with the results plotted and concentration calculated using our developed GUI. The quantification of droplet single-cell sequencing results for various barcodes has been accomplished by analyzing 2,000 droplets (across 10 images) within a remarkable 2.5 seconds. This approach demonstrates a significant improvement in speed and accuracy compared to conventional methods that rely on manual counting. To the best of our knowledge, this is the first successful implementation of a large-scale, multi-functional, and all-in-one image analysis algorithm for single-cell sequencing and numerous other biological experiments.

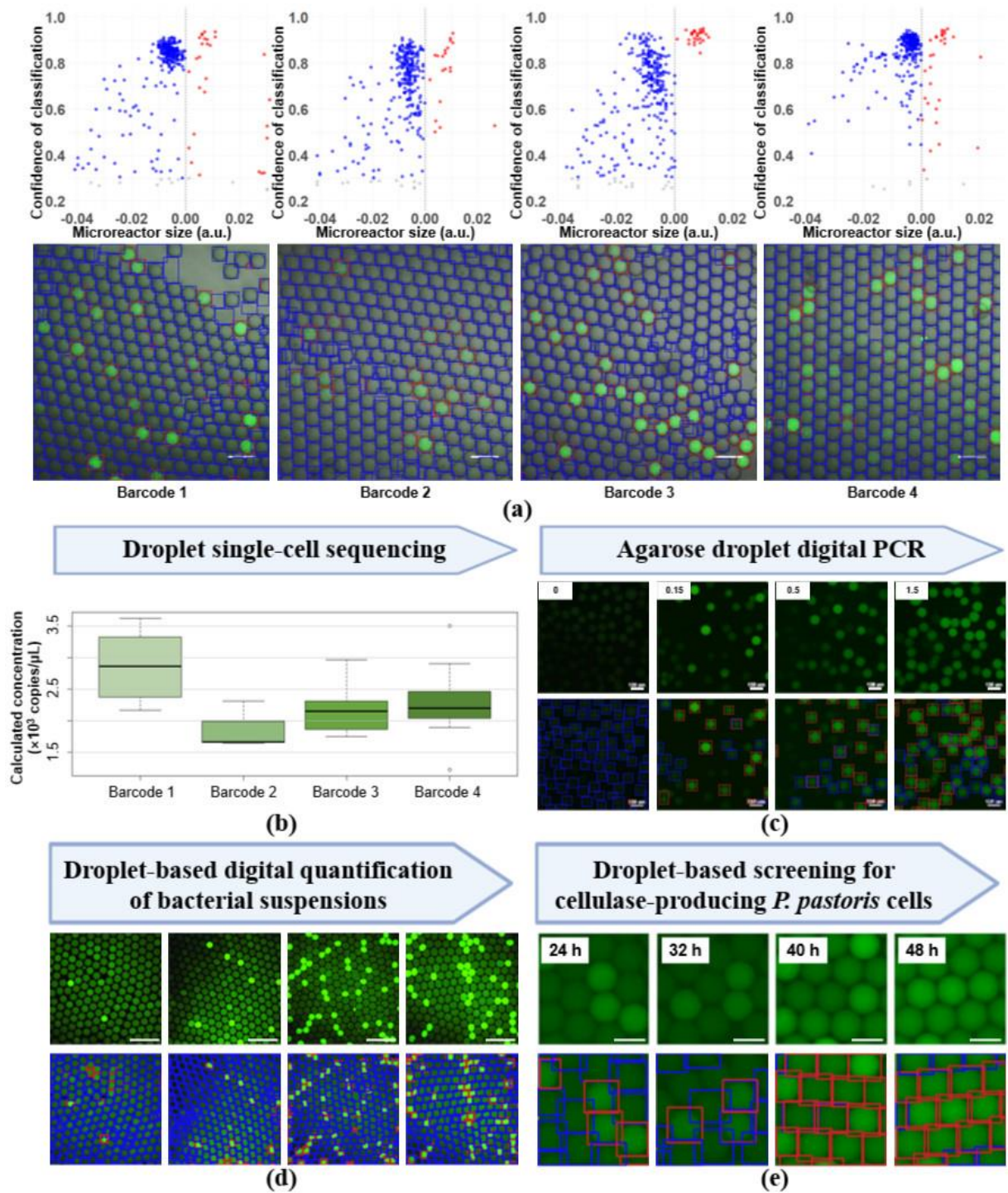

**Fig. 6 Validation of Deep-dGFP generalization capability: microreactors segmentation and classification results for various biological applications.** Our Deep-dGFP algorithm shows strong generalization capability for other different GFP-labeled microreactors biological applications, including droplet-based, microwell-based, and agarose-based experiments. (a) Droplet segmentation and classification results for droplet single-cell sequencing of different barcoding. (Scale bar = 100 μm). (b) Validation of the above analysis results has been conducted

through statistical analysis. For each experiment condition, more than 2,000 droplets were analyzed, with inferenced concentration plotted correspondingly. (c) Image analysis results of representative frames for agarose-based digital PCR of different encapsulation rates. (Scale bar = 100 μm). (d) Image analysis results of representative frames for droplet-based digital bacterial quantification under different suspension conditions. (Scale bar = 100 μm). (e) Image analysis results of representative frames for droplet-based screening for cellulase-producing *Pichia pastoris* cells. (Scale bar = 50 μm).

### 3. Discussion and Conclusion

The performance of our deep-dGFP detection pipeline, which captures GFP-labeled images in a two-dimensional area, is influenced by multiple factors, with input image quality being the most crucial. It is governed by factors such as focusing, resolution, luminance, and saturation. To enhance its performance, we optimized the settings of fluorescence imaging to ensure sharpness in the captured images, based on the detection results obtained from the collected images. This optimization process improves both the throughput and accuracy of the algorithm.

In addition to the aforementioned parameters, the performance of the trained deep-learning algorithm and the processing speed of the controlling laptop that reads images and displays plots can also be a limiting factor. The data labeling process involved 206 frames of nucleic acid concentrations ranging from 0 to 5 ng/μL. Data augmentation was performed by enhancing contrast by 20-30%, applying a noise ratio of 3%, and blurring images by 5-20%. The original images and labeled data were then divided into three sets: training data (80%), validation data (10%), and testing data (10%). No operations were applied to the collected images for testing. Our technique exhibits robustness due to the diverse input training data, including focusing and defocusing images, different exposure conditions, droplets of various sizes (18 to 114 pixels), image sizes ranging from 256, 256, 3 to 900, 900, 3 pixels, and magnifications from 4× to 10×. This makes it easily transferable to different GFP application scenarios, such as quantification for environmental toxicity.

Currently, our device operates in two modes: offline mode and real-time mode, depending on the user's requirements. In the offline mode, we collect images and videos using lab computers and store the original materials in a designated folder. As mentioned in the Results section, our trained algorithm is capable to analyze both fluorescence field images and merged images. After completing the experiment, we perform image analysis by retrieving the images from the folders, which also allows us to fine-tune parameters such as linewidth of labeling bounding-boxes for optimization purposes. In terms of runtime analysis, the data analysis and display process take approximately 0.274 seconds per field of view (FOV) frame in offline mode, allowing for the automatic segmentation, classification, and plotting of up to 10,000 droplets simultaneously. Additionally, our technique enables analyzing and displaying GFP-labeled samples in a real-time manner. In real-time mode, we capture and display one frame per second. The main computational operations involved are as follows: (1) reading images from the API or SDK cable of the fluorescence microscope, taking around 0.033 seconds per image; (2) utilizing the Yolov5 algorithm for image analysis, including droplet segmentation, classification, and measurement, which takes approximately 0.194 seconds per image; (3) outputting and saving labeled images and analyzed results (including class, confidence, coordinates, and microreactor size) of each droplet into a designated folder, requiring around 0.004 seconds per image; (4) displaying the results on the designed GUI, which takes approximately 0.541 seconds per image. In summary, our technique ensures automatic image collection, analysis, display, and plotting with a delay of less than 1.772 seconds.

In our experiments, since the input droplet size of training data varies from 18 pixels to 114 pixels, fluorescence noise due to potential fouling or specks of dust is non-inferenced. Additionally, the background signal exhibited negligible spatial variation, and the exposure was carefully adjusted during the experiments, allowing for a certain tolerance of light transmission loss. Moreover, since our deep-dGFP pipeline employs disposable microfluidic chips that can be easily replaced, routine evaluation of the background signal can be implemented to automatically alert users when a chip replacement is necessary. To specify the limitations of our technique, we have determined that our method consistently delivers stable performance for input microreactor sizes up to 1/57 of the total image pixels, while the deep-learning algorithm fails when the image size exceeds 3096 pixels (**Supplementary Fig. 3**). This limitation arises from the fact that the input images are zoomed to 640, 640, 3 pixels before analysis, rendering microreactors beyond the specified size limit unanalyzable. However, this issue can be resolved by cropping the image into several smaller images, ensuring that the size of each cropped region corresponds to approximately 1/32 of the total region of interest (ROI).

In conclusion, the Deep-dGFP pipeline is a cost-effective and automated analysis tool that represents a significant advancement in the field of biological quantification. This is achieved through the integration of accessible hardware, including a microfluidics chip, thermal cycler, and camera-coupled fluorescence microscope, with a state-of-the-art deep learning algorithm. The quantification of over 2,000 microreactors (across 10 images) takes remarkably 2.5 seconds. GFP, as a reporter gene, is much less harmful than FITC, making it widely applicable in biological experiments, live cells, and even live organisms. We anticipate that the Deep-dGFP technique will be readily adopted by biomedical laboratories and has the potential to be developed into a point-of-care digital nucleic acid test system. Furthermore, its applicability can be expanded by utilizing other color mutant fluorescence probes such as blue fluorescent protein (EBFP, EBFP2, Azurite, mKalama1), cyan fluorescent protein (ECFP, Cerulean, CyPet, mTurquoise2), and yellow fluorescent protein derivatives (YFP, Citrine, Venus, YPet).

## 4. Materials and methods
### 4.1 Sample preparation

The primary commercial kit utilized for the ddPCR reactions was the Q5® High-Fidelity DNA Polymerase (New England Biolabs). Diluted restriction enzymes, dNTPs, buffer, forward and reverse primers, Tween-20, PEG-8000, and PCR water were added. Details regarding the composition of reagents utilized in the ddPCR reaction are available in **Supplementary Table 1**. In the ddPCR experiments, the cDNA templates were serially diluted. The stock cDNA concentration was $(1.12 \pm 0.09) \times 10^4$ copies $\mu L^{-1}$.

In the ddPCR experiments shown in **Fig.3**, the PCR mixture was prepared with 1X KAPA HIFI buffer, 0.3 mM dNTP, 1X KAPA HIFI polymerase, 0.3 uM forward and reverse primers, templates, 0.1% NP-40, 0.2% Tween 20, and 0.1 mg/mL BSA (NEB, USA). The template concentrations were 40 pg, 4 pg, and 0.4 pg per 20 uL PCR system, respectively. The PCR was conducted with the following procedures: 95 degrees for 3 min; 45 cycles of 98 degrees for 20 s, 61 degrees for 15 s, and 72 degrees for 15s; 72 degrees for 1 min and 12 degrees for infinite holding. The PCR template was the extracted seahorse (*Hippocampus kuda*) genome and the targeted region was located at the *cytochrome c oxidase subunit I (COI)* with an amplicon size of 206 bp; the primers are as following:

Forward primer (5→3): TTTCTTCTCCTCCTTGCTTCCTCAG

Reverse primer (5→3): GAAATTGATGGGGGTTTTATGTTG

In the ddPCR experiments shown in **Fig.4**, the PCR mixture was prepared with 1X Platinum SuperFi II buffer, 0.2 mM dNTP, 1X Platinum SuperFi II polymerase, 0.5 uM forward and reverse primers, templates, 0.2% Tween 20, 0.2 mg/mL BSA (NEB, USA), and 0.4% PEG8000. The PCR was conducted with the following procedures: 98 degrees for 30 s; 25 or 45 cycles of 98 degrees for 10 s, 60 degrees for 10 s, and 72 degrees for 15s; 72 degrees for 5 min final extension and 12 degrees for infinite holding. Designed sequences of templates and primers are as following:

The PCR template (5→3):

GTCTCGTGGAGCTCGACAGCATNNNNNNTGNNNNNNTGCCTACGACAAACAGACCT AAAATCGCTCATTGCATACTCTTCAATCAG

Forward primer (5→3):

Acrydite-ACTAACAATAAGCTCUAUAGTCTCGTGGAGCTCGACAG

Reverse primer (5→3):

CTGATTGAAGAGTATGCAATGAG

In the single-cell sequencing experiments shown in **Fig.6**, the PCR mixture was prepared with 1X TaKaRa PrimeSTAR GXL buffer, 0.2 mM dNTP, 1X TaKaRa PrimeSTAR GXL polymerase, 0.2 uM forward and reverse primers, templates, 0.5% Tween 20, 0.1 mg/mL BSA (NEB, USA), and 0.5% PEG8000. The PCR program was set as follows: 25 cycles at 95ºC 10 s, 59ºC 15 s and 68ºC 15 s followed by a final extension at 68ºC 3-min and a hold at 25ºC. Sequences of templates and primers are shown in Tabel 1.

Table 1. Sequences in single-cell sequencing experiments of **Fig. 6**

| | | | Sequence (5→3) |
|---|---|---|---|
| Barcode 1 | Template | | GTCTCGTGGAGCTCGACAGNNNNNNNNNNNN TCGCTCATTGCATACTCTTCAATCAGC |
| | Primer | Forward | GTCTCGTGGAGCTCGACAG |
| | | Reverse | GCTGATTGAAGAGTATGCAATG |
| Barcode 2 | Template | | GTCTCGTGAGTCAGGACAGNNNNNNNNNNNN TCGCTCATTGCATACTCTTCAATCAGC |
| | Primer | Forward | GTCTCGTGAGTCAGGACAG |
| | | Reverse | GCTGATTGAAGAGTATGCAATG |
| Barcode 3 | Template | | GTCTCGTGACCTCGGACAGNNNNNNNNNNN |

|           |          |         | TCGCTCATTGCATACTCTTCAATCAGC |
|-----------|----------|---------|------------------------------|
|           | Primer   | Forward | GTCTCGTGACCTCGGACAG |
|           |          | Reverse | GCTGATTGAAGAGTATGCAATG |
| Barcode 4 | Template |         | GTCTCGTGGACAGTGACAGNNNNNNNNNNNN |
|           |          |         | TCGCTCATTGCATACTCTTCAATCAGC |
|           | Primer   | Forward | GTCTCGTGGACAGTGACAG |
|           |          | Reverse | GCTGATTGAAGAGTATGCAATG |

### 4.2 Microfluidic chip fabrication and droplet generation

A flow-focusing microfluidic chip (as depicted in **Supplementary Fig. 2**) possessing a cross-sectional dimension of 30.0 μm in width and 38.5 μm in height was designed and fabricated to generate the droplets. The chip was fabricated utilizing a standard process of SU-8 photolithography and PDMS replica molding. An initial master mold of 60.0 μm height was prepared by a spinning photoresist (SU8-2075) at 5000 r.p.m for 28 s on a 4" silicon wafer, followed by UV exposure, baking, and developer bath according to the manufacturer's specifications. Next, a base-to-curing ratio of 10:1 of PDMS prepolymers (Dow, Inc., Sylgard 184) was degassed, poured onto SU-8 masters, and baked at 60 °C for 10 h before being removed and cut to the desired shape. Inlets and outlets were then punched through the PDMS, and oxygen plasma treatment was employed to treat both the PDMS slabs and glass slides for 1 min, after which they were bonded by brief baking at 110 °C. Finally, the microfluidic devices were hydrophobized by baking at 55 °C for 24 h.

Monodisperse emulsions were generated using the homemade microfluidic, whereby the flow was driven by two syringe pumps (Legato 100, KD Scientific or Ph.D. 2000, Harvard Apparatus, USA) at the inlets. Emulsion generation utilized commercial droplet generation oil (1864006, Bio-Rad Laboratories, Inc.) as the continuous phase of the emulsions. The resultant droplets exhibited a uniform size distribution with a mean diameter of 46.37 ± 1.64 μm (0.052 nL), at flow rates of 100 μL/min for the water phase (including ddPCR mix and cDNA samples, 25 μL in total) and 100 μL/min for the oil phase (droplet generation oil, 100 μL, Bio-Rad). The generated droplets were collected and transferred into 0.2 ml PCR stripe tubes, which were covered with mineral oil to minimize evaporation. The samples were subjected to polymerase chain reaction (PCR) reactions in a thermal cycler (Bio-Rad) with a hot start at 94 °C for 5 min, followed by 55 cycles of denaturation at 94 °C for 30 s, annealing and extension at 60 °C for 1 min. Subsequently, the droplets were isolated onto a glass slide for observation.

### 4.3 Microreactors imaging and image processing

The amplified droplets were collected and dispensed into a specially designed Polydimethylsiloxane (PDMS) chamber for observation under a lab fluorescence microscope. Imaging of the emulsions was conducted using an inverted microscope (Eclipse Ti-U, Nikon) coupled with a camera (DS-Qi2, Nikon) in both brightfield and fluorescence fields. Fluorescence excitation was achieved at 455 nm, and the emitted light was captured by a CCD through a 495 nm long-pass filter. Images were acquired at 4× magnification with an exposure time of 1 s and a sensor sensitivity of 200. For each sample, 10 images were taken at different positions excluding edges, including over 2,000 droplets in total. Video recording was performed at a frame rate of 20 frames per second, with each video lasting between 3 and 5 minutes, resulting in approximately 5,400–9,000 frames. The fluorescence intensity of classified droplets was quantified by determining the mean grayscale using ImageJ and normalizing it to the grayscale range of the recorded images.

The sequential diluted microwell dPCR original images in **Fig. 5** were supported by collaborator Prof. Mingli You (School of Life Science and Technology, Xi'an Jiaotong University) [23]. The agarose digital PCR original images in **Fig. 6 (c)** were provided by collaborators Dr. Xuefei Leng and Prof. Chaoyong Yang (College of Chemistry and Chemical Engineering, Xiamen University) [24]. The original images in **Fig. 6 (d)** and **(e)** were captured from reported journal articles [25] and [26], respectively.

### 4.4 Deep-learning-enabled automatic data analysis

During the detection process, data analysis was conducted using both frames and videos. The labeling and training processes included (1) Image labeling, (2) Data augmentation, (3) Data splitting, (4) Detection algorithm (Yolo-V5m) training, and (5) Detection results. For the results obtained with deep-learning assistance, training data were generated through the utilization of Roboflow for online annotation and model training. The training region of interest (ROI) per image was set to 512, with the maximum number of ground truth instances per image being limited to 100. To generate a sufficient number of proposals, the non-maximum suppression threshold was defined as 0.95. The anchor box sizes were set to 32, 64, 128, 256, and 512 pixels, with strides being 4, 8, 16, 32, and 64 pixels and width-length ratios of 0.5, 1, and 2. The maximum number of instances per image was set to 10,000 during the detection process. The assessment of the performance of ddPCR encompassed an evaluation of its linearity range, limit of quantification (LOQ), and reproducibility.

Sample concentration inference by fitting into Poisson distribution was determined by employing Equation (8). For each concentration, over 2,000 microreactors (including droplets and microwells) were analyzed.

The probability $\Pr(X = k)$ that a microreactor will contain k copies of target gene if the mean number of target copies per microreactor is $\lambda$:

$$f(k, \lambda) = \Pr(X = k) = \frac{\lambda^k e^{-\lambda}}{k!} \tag{1}$$

where

$k$ is the number of occurrences ($k$ can take values 0, 1, 2, ...).

$e$ is Euler's number ($e = 2.71828…$).

! is the factorial function.

Inputting $k=0$ gives the probability that a microreactor will be empty:

$$\Pr(X = 0) = e^{-\lambda} \tag{2}$$

For the number of microreactors being large enough, the observed fraction of empty microreactors (E) gives estimation of $Pr(X = 0)$

$$E = e^{-\lambda} \tag{3}$$

At the same time, by definition of E,

$$E = \frac{N_{negative}}{N} \tag{4}$$

Solving (3) we get

$$\lambda = -\ln(E) \tag{5}$$

As $\lambda$ is the copies per microreactor, concentration of copies per volume is

$$Concentration = \frac{\lambda}{V_{microreactor}} \tag{6}$$

Which means

$$Concentration = \frac{-\ln(E)}{V_{microreactor}} \tag{7}$$

Combining (4) and (7), we get

$$Concentration = -\ln\left(\frac{N_{negative}}{N_{total}}\right)/V_{microreactor} \tag{8}$$

### 4.5 Graphical user interface

We have developed a versatile GUI that enables real-time display of analyzed results generated by the algorithm. This GUI allows for visual inspection of GFP-labeled microreactors using laboratory microscopes. It provides a display of original images captured at a rate of 1 frame per second, with droplets labeled according to their class and confidence. Additionally, the GUI plots the droplets by accumulating results from sequential frames. Users have the option to digitally save the raw images, background-subtracted images, plot results, and calculated values. The size of the droplets and the calculated template concentration are continuously displayed at the bottom. Furthermore, the GUI can be operated in off-line mode to analyze pre-saved image datasets by reading folders. The design codes for the GUI can be found in the Supporting Information.

### 4.6 Statistical Analysis

The statistical analysis was carried out using GraphPad Prism software (GraphPad Software). In all cases, data represent mean ± standard deviation (SD) with n ≥ 3. A t-test was adopted for hypothesis testing, and significance was defined as $p \leq 0.05$.

## 5. References

[1] "Measurement of Nucleic Acid ConcentrationsUsing the DyNA Quant☐☐and the GeneQuant☐."

[2]    R. Brankatschk, N. Bodenhausen, J. Zeyer, and H. Burgmann, "Simple absolute quantification method correcting for quantitative PCR efficiency variations for microbial community samples," *Appl. Environ. Microbiol.*, vol. 78, no. 12, pp. 4481–4489, Jun. 2012.

[3]    Y. Wei, G. Cheng, H. P. Ho, Y. P. Ho, and K. T. Yong, "Thermodynamic perspectives on liquid-liquid droplet reactors for biochemical applications," *Chem. Soc. Rev.*, vol. 49, no. 18, pp. 6555–6567, 2020.

[4]    M. Postel, A. Roosen, P. Laurent-Puig, V. Taly, and S. F. Wang-Renault, "Droplet-based digital PCR and next generation sequencing for monitoring circulating tumor DNA: a cancer diagnostic perspective," *Expert Rev. Mol. Diagn.*, vol. 18, no. 1, pp. 7–17, 2018.

[5]    V. Y. Cusenza, A. Bisagni, M. Rinaldini, C. Cattani, and R. Frazzi, "Copy number variation and rearrangements assessment in cancer: Comparison of droplet digital pcr with the current approaches," *Int. J. Mol. Sci.*, vol. 22, no. 9, 2021.

[6]    I. Palacín-aliana, N. García-romero, A. Asensi-puig, J. Carrión-navarro, V. González-rumayor, and Á. Ayuso-sacido, "Clinical utility of liquid biopsy-based actionable mutations detected via ddpcr," *Biomedicines*, vol. 9, no. 8, pp. 1–20, 2021.

[7]    B. Chen, Y. Jiang, X. Cao, C. Liu, N. Zhang, and D. Shi, "Droplet digital PCR as an emerging tool in detecting pathogens nucleic acids in infectious diseases," *Clin. Chim. Acta*, vol. 517, no. March, pp. 156–161, 2021.

[8]    L. B. Pinheiro *et al.*, "Evaluation of a droplet digital polymerase chain reaction format for DNA copy number quantification," *Anal. Chem.*, vol. 84, no. 2, pp. 1003–1011, 2012.

[9]    A. S. Whale, J. F. Huggett, and S. Tzonev, "Fundamentals of multiplexing with digital PCR," *Biomolecular Detection and Quantification*, vol. 10. Elsevier GmbH, pp. 15–23, 01-Dec-2016.

[10]   B. T. Lau, C. Wood-Bouwens, and H. P. Ji, "Robust Multiplexed Clustering and Denoising of Digital PCR Assays by Data Gridding," *Anal. Chem.*, vol. 89, no. 22, pp. 11913–11917, Nov. 2017.

[11]   W. Zhou *et al.*, "Low Cost, Easily-Assembled Centrifugal Buoyancy-Based Emulsification and Digital PCR," *Micromachines*, vol. 13, no. 2, 2022.

[12]   B. Demaree, D. Weisgerber, A. Dolatmoradi, M. Hatori, and A. R. Abate, "Direct quantification of EGFR variant allele frequency in cell-free DNA using a microfluidic-free digital droplet PCR assay," *Methods Cell Biol.*, vol. 148, pp. 119–131, 2018.

[13]   S. Bartkova, M. Vendelin, I. Sanka, P. Pata, and O. Scheler, "Droplet image analysis with user-friendly freeware CellProfiler," *Anal. Methods*, vol. 12, no. 17, pp. 2287–2294, 2020.

[14]   J. Chen *et al.*, "Capillary-based integrated digital PCR in picoliter droplets," *Lab Chip*, vol. 18, no. 3, pp. 412–421, 2018.

[15]   F. Cui, Y. Yue, Y. Zhang, Z. Zhang, and H. S. Zhou, "Advancing Biosensors with Machine Learning," *ACS Sensors*, vol. 5, no. 11, pp. 3346–3364, 2020.

[16] K. A. Vakilian and J. Massah, "A Portable Nitrate Biosensing Device Using Electrochemistry and Spectroscopy," *IEEE Sens. J.*, vol. 18, no. 8, pp. 3080–3089, 2018.

[17] A. Morellos *et al.*, "Machine learning based prediction of soil total nitrogen, organic carbon and moisture content by using VIS-NIR spectroscopy," *Biosyst. Eng.*, vol. 152, pp. 104–116, 2016.

[18] Z. Hu *et al.*, "A novel method based on a Mask R-CNN model for processing dPCR images," *Anal. Methods*, vol. 11, no. 27, pp. 3410–3418, 2019.

[19] H. Yang *et al.*, "A deep learning based method for automatic analysis of high-throughput droplet digital PCR images," *Analyst*, vol. 148, no. 2, pp. 239–247, Nov. 2022.

[20] M. Zimmer, "Green fluorescent protein (GFP): Applications, structure, and related photophysical behavior," *Chem. Rev.*, vol. 102, no. 3, pp. 759–781, Mar. 2002.

[21] •, "Using GFP to see the light."

[22] M. Chalfie, "GREEN FLUORESCENT PROTEIN," *Photochem. Photobiol.*, vol. 62, no. 4, pp. 651–656, 1995.

[23] C. Cao *et al.*, "Similar color analysis based on deep learning (SCAD) for multiplex digital PCR via a single fluorescent channel," *Lab Chip*, vol. 22, no. 20, pp. 3837–3847, Sep. 2022.

[24] X. Leng, W. Zhang, C. Wang, L. Cui, and C. J. Yang, "Agarose droplet microfluidics for highly parallel and efficient single molecule emulsion PCR," *Lab Chip*, vol. 10, no. 21, pp. 2841–2843, Nov. 2010.

[25] É. Geersens, S. Vuilleumier, and M. Ryckelynck, "Growth-Associated Droplet Shrinkage for Bacterial Quantification, Growth Monitoring, and Separation by Ultrahigh-Throughput Microfluidics," *ACS Omega*, vol. 7, no. 14, pp. 12039–12047, Apr. 2022.

[26] H. Yuan *et al.*, "Microfluidic screening and genomic mutation identification for enhancing cellulase production in Pichia pastoris," *Biotechnol. Biofuels Bioprod.*, vol. 15, no. 1, Dec. 2022.

## 6. Acknowledgements

The authors are grateful to the funding support from the Hong Kong Research Grants Council (project reference: GRF14204621, GRF14207920, GRF14207419, GRF14203919, N_CUHK407/16), and the Innovation and Technology Commission (project reference: GHX-004-18SZ).

The authors would like to acknowledge Prof. Mingli You (School of Life Science and Technology, Xi'an Jiaotong University), Dr. Xuefei Leng and Prof. Chaoyong Yang (College of Chemistry and Chemical Engineering, Xiamen University), Ms. Khadija BIBI, Dr. Xiaoyu Ji, Dr. Shiyue Liu, Dr. Guangyao Cheng, Dr. Shutian Zhao, Dr. Md Habibur Rahman, Mr. Shirui Zhao, and Ms. Syeda Aimen Abbasi (Department of Biomedical Engineering, The Chinese University of Hong Kong) for their support in the project development.

## 7. Author information

Authors and Affiliations

**Department of Biomedical Engineering, The Chinese University of Hong Kong, Shatin, Hong Kong SAR, 999077, China**

Yuanyuan Wei, Sai Mu Dalike Abaxi, Nawaz Mehmood, Luoquan Li, Fuyang Qu, Guangyao Cheng, Dehua Hu, Yi-Ping Ho, Wu Yuan & Ho-Pui Ho

**Centre for Biomaterials, The Chinese University of Hong Kong, Hong Kong SAR, 999077, China**

Yi-Ping Ho

**Hong Kong Branch of CAS Center for Excellence in Animal Evolution and Genetics, Hong Kong SAR, 999077, China**

Yi-Ping Ho

**The Ministry of Education Key Laboratory of Regeneration Medicine, Hong Kong SAR, 999077, China**

Yi-Ping Ho

Contributions

Y. Wei and N. Mehmood contributed to the study's conception and design, with N. Mehmood specifically developing the deep-learning algorithm training. S. Abaxi developed the calculation function and real-time display GUI. Y. Wei was responsible for data labeling. Y. Wei and F. Qu conducted the biological experiments. Y. Wei performed the image analysis experiments. Y. Wei and D. Hu and Y. Wei performed data analysis and figures editing. F. Qu, L. Li, G. Cheng, and Y. Ho supplied the microfluidic chip and customized microfluidic platform. Y. Wei and S. Abaxi conducted on-site experiments. Y. Wei wrote the manuscript with contributions from all authors. W. Yuan and H. Ho conceived the project and supervised the research.

Corresponding author

Correspondence to Scott Wu-Yuan and Ho-Pui Ho.

## 8. Ethics declarations

No conflict of interest

## 9. Additional information

Electronic supplementary material

Supplementary Information